\documentclass[twocolumn,showpacs,preprintnumbers,amsmath,amssymb]{revtex4}
\usepackage{epsfig}

\def\be{\begin{equation}}
\def\ee{\end{equation}}
\def\beq{\begin{equation}}
\def\eeq{\end{equation}}
\def\bea{\begin{eqnarray}}
\def\eea{\end{eqnarray}}
\def\mc{\mathcal}

\begin{document}

\preprint{DAMTP-2008-5}
\preprint{CAVENDISH-HEP-2008-02}
\title{Measuring Smuon-Selectron Mass Splitting
at the LHC
and Patterns of
Supersymmetry Breaking}

\author{B.C.  Allanach$^{1}$}
\author{J.P. Conlon$^{1,2}$}
\author{C.G. Lester$^{2}$}
\affiliation{$^{1}$ DAMTP, Centre for Mathematical Sciences, Wilberforce Road,
University of Cambridge, 
Cambridge, CB3 0FY, United Kingdom}
\affiliation{$^{2}$Cavendish Laboratory, University of Cambridge, JJ Thomson
  Avenue, Cambridge, CB3 0HE, United Kingdom}

\begin{abstract}
With sufficient data, Large Hadron Collider (LHC) experiments can constrain
the selectron-smuon mass 
splitting
through differences in the di-electron and di-muon edges from supersymmetry
(SUSY) cascade
decays. We study the sensitivity of the LHC to this mass splitting, which 
within mSUGRA may be
constrained down to $\mc{O}(10^{-4})$ for 30 fb$^{-1}$ of integrated
luminosity. 
Over substantial regions of SUSY breaking parameter space the
fractional edge splitting can be significantly enhanced over the fractional
mass splitting.
Within models where the selectron and smuon are constrained to be universal at
a high scale, edge splittings up to a few percent may be induced
by renormalisation group effects and may be significantly discriminated from zero. 
The edge splitting provides important information about high-scale SUSY
breaking terms and should be included in any fit of LHC data
to high-scale models.
\end{abstract}


\maketitle
TeV-scale supersymmetry is one of the most promising solutions to the weak
hierarchy problem of the
Standard Model. If TeV-scale supersymmetry is realized in nature, the LHC
will be able to discover it by the production of sparticles and
measure some of the parameters of the minimal supersymmetric standard model (MSSM).
Through the use of renormalisation group equations (RGEs) to
connect the low and high scale 
Lagrangians, the measured MSSM parameters can be used to test hypotheses, such
as flavor universality (or non-universality) of the high-scale soft SUSY
breaking terms. Such a link
would be a welcome additional empirical window into the unsolved problem of
flavor and how it relates to SUSY breaking.

Potentially one of the most accurate LHC measurements of SUSY is that of the
di-lepton edge arising 
from neutralino-slepton-neutralino SUSY cascade decays,
$$
{\chi}_2^0 \to \tilde{l}^{\pm}l^{\mp} \to \chi_1^0 l^{\pm} l^{\mp}.
$$
If the mass ordering $m_{\chi_2^0} > m_{\tilde{l}} > m_{\chi_1^0}$ is present,
the di-lepton mass spectrum has a prominent kinematic edge at
\begin{equation}
m_{ll}^2 =  \frac{(m_{{\chi}_2^0}^2 - m_{\tilde{l}}^2)(
m_{\tilde{l}}^2 - m_{{\chi}_1^0}^2)}{m_{\tilde{l}}^2}. \label{eq:edge}
\end{equation}
As long as $\vert m_{{\chi}_{1,2}^0} - m_{\tilde{\mu}} \vert \gg m_{\mu}$
the finiteness of the muon mass induces fractional
corrections that are $\mc{O}\left( m_\mu^2/m_{susy}^2 \right)$ (where
$m_{susy}$ denotes the order of magnitude of sparticle masses, $\gtrsim 100$ GeV) and
may be 
neglected, as they are throughout this paper.
Such an edge may be measured at LHC experiments 
with per-mille precision~\cite{Armstrong:1994it}.
Although the original authors commented that one could measure the di-lepton
and di-muon endpoints separately \cite{hepph9907518},
it is generally assumed, motivated by the MSSM flavor problem, that $m_{\tilde{e}} = m_{\tilde{\mu}}$, with the
$m_{ee}$ and
$m_{\mu \mu}$ edges occurring at identical values.

The flavor problem is
one of the most pressing questions of supersymmetric phenomenology.
The Lagrangian of a TeV-scale MSSM is highly constrained by the absence of new
contributions to
low energy flavor changing neutral currents (FCNCs).
Classic examples are $K^0 \bar{K}^0$ mixing or
in the leptonic sector
the $\mu \to e \gamma$ branching ratio.
These strongly suggest that new physics present at the TeV scale should obey to high
accuracy the principle of MFV (Minimal Flavor Violation), at least in the first two
generations.
The $\mu \to e \gamma$ branching ratio, which is currently bounded at $BR(\mu \to e \gamma) < 1.2 \times 10^{-11}$ \cite{hepex9905013}
and in the future will be measured to at least $10^{-13}$ \cite{PSI},
constrains off-diagonal propagator mixing between the smuon
and selectron $LL$ flavor eigenstates to \cite{Ciuchini:2007ha},
$$
\delta_{12,LL} \lesssim 6 \times 10^{-4}.
$$
The $RR$ constraints are similar over most of parameter space, but potential
cancellations exist and so $\delta_{12,RR}$ has a weaker bound than
$\delta_{12,LL}$.

Nonetheless, exact low-scale flavor universality is not expected; some non-universality
will be automatically induced by RGEs and in the context of string theory models of supersymmetry breaking
loop effects are expected to violate flavor universality even if 
leading order physics preserves it \cite{07100873}.
It is therefore important to investigate the sensitivity of the LHC to the mass
splitting between the right-handed selectron and smuon:
\begin{equation}
\Delta m^2 \equiv m_{{\tilde \mu}_R}^2 - m_{{\tilde e}_R}^2.
\end{equation}
$\Delta m^2$ is not directly constrained by the $BR(\mu \to e \gamma)$
measurement 
as it does not give lepton flavor violation (LFV).
LFV effects at the LHC have been
studied in various papers
\cite{hepph9904422,
hepph0010086,hepph0202129,hepph0206148,hepph0510074,07061845, 07120674}
but are not our focus here.

We may relate $\Delta m^2$ to the SUSY breaking terms defined at a high
scale $M_X$ through the MSSM RGEs. At one-loop
order\cite{Martin:1993zk},
 \bea
   16 \pi^2 \frac{d \Delta m^2}{d \ln \mu} & = & 4 \Bigg[
 Y_\mu^2 (m_{{\tilde \mu}_R}^2 + m_{{\tilde \mu}_L}^2) -
 Y_e^2 (m_{{\tilde e}_R}^2 + m_{{\tilde e}_L}^2) +  \nonumber \\
 & & m_{H1}^2 (Y_\mu^2 - Y_e^2)+ Y_\mu^2 h_\mu^2 - Y_e^2 h_e^2 \Bigg],
 \label{dmrge}
 \eea
where $\mu$ is the $\overline{DR}$ renormalisation scale, and $m_{{\tilde
    \mu}_R}$, $m_{{\tilde \mu}_L})$, $m_{{\tilde
    e}_R}$, $m_{{\tilde e}_L}$ and $m_{H1}$ are the soft SUSY breaking
masses for
right- and left-handed smuons and selectrons and the Higgs field,
evaluated at a renormalisation scale $\mu$. $h_e,h_\mu$
are the trilinear selectron and smuon soft 
SUSY breaking couplings evaluated at a renormalisation scale $\mu$.
At $M_Z$, we have the boundary condition $Y_\mu(M_Z) = \sqrt{2} m_\mu(M_Z) /
(v \cos \beta)$, where $v=\sqrt{v_1^2(M_Z) + v_2^2(M_Z)}$, $v_{1,2}$ being the MSSM
Higgs vacuum expectation values. We may solve the RGE for $Y_\mu$,
$Y_\mu(M_X) = Y_\mu(M_Z) + \mathcal{O}( \ln(M_X/M_Z) / 16 \pi^2 )$.
We neglect
the ratio of the electron Yukawa coupling to the muon Yukawa
coupling ($Y_e/Y_\mu$) and
solve eq.~(\ref{dmrge}) to first order in $\ln
(M_X/M_Z)/(4 \pi)^2$ to obtain
\begin{eqnarray}
\Delta m^2(M_Z)  &=&  \Delta m^2(M_X) + \frac{8 m_\mu^2}{16 \pi^2 v^2} \Bigg[
m_{{\tilde \mu}_R}^2(M_X)  \nonumber \\
&& + m_{{\tilde \mu}_L}^2(M_X)
  + m_{H_1}^2 (M_X) + \nonumber \\
&&A_\mu^2(M_X)
\Bigg]  \tan^2 \beta \ln \left( \frac{M_X}{M_Z} \right), \label{leadinglog}
\end{eqnarray}
where $\tan \beta=v_2/v_1$.
We have allowed for the presence of a primordial mass splitting $\Delta
m^2(M_X)$ and used the large $\tan \beta$ limit in eq.~(\ref{leadinglog}) so
that $\cos \beta \approx
1/\tan \beta$. We have also neglected the QED running of the muon mass below
the electroweak scale.
We see from eq.~(\ref{leadinglog}) that the magnitude of
$\Delta m^2$ is always enhanced as $\tan^2 \beta \ln
(M_X/M_Z)$. The evaluation of $\Delta m^2(m_Z)$ is complicated by the unknown boundary
conditions on $m_{{\tilde \mu}_R}^2$, $m_{{\tilde \mu}_L}^2$,
$m_{H_1}^2 $ and $A_\mu^2$ at $M_X$, but within constrained models other edge
measurements are expected to help pin them down. For example, in the mSUGRA
model, the SUSY breaking scalar masses have a universal value of $m_0$ at $M_X$ and
the trilinear SUSY breaking scalar couplings have a universal value of
$A_0$. In this case, $\Delta m^2(M_X)=0$  and
the quantity in square brackets is simply $3 m_0^2 + A_0^2$, which
may be bounded from other measurements.

From eq.~\ref{eq:edge}
the variation of the edge position with the slepton mass is given by
$$
\frac{d m^2_{ll}}{dm_{\tilde{l}}^2} = \frac{m_{\chi_1^0}^2 m_{\chi_2^0}^2}{m_{\tilde{l}}^4} - 1,
$$
with a fractional shift in the invariant mass edge of
\be
\frac{\Delta m_{ll}}{m_{ll}} = \frac{\Delta m_{\tilde{l}}}{m_{\tilde{l}}}
\left( \frac{m_{\chi_1^0}^2 m_{\chi_2^0}^2 - m_{\tilde{l}}^4}{(m_{\chi_2^0}^2 - m_{\tilde{l}}^2 )
(m_{\tilde{l}}^2 - m_{\chi_1^0}^2)} \right), \label{edgesplitting}
\ee
up to terms of $\mc{O}\left(\frac{\Delta
  m_{\tilde{l}}}{m_{\tilde{l}}}\right)^2$.

\begin{figure}
\epsfig{file=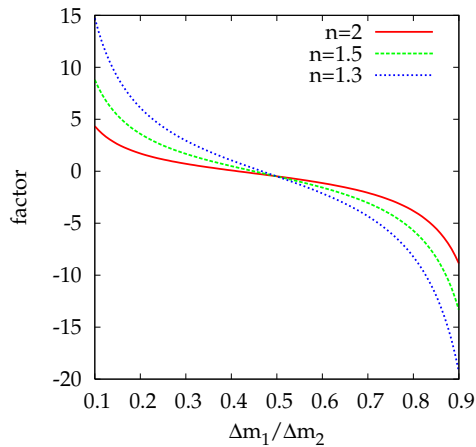, width=\columnwidth}
\caption{The enhancement factor $(\Delta m_{ll}/m_{ll})/(\Delta
  m_{\tilde{l}}/m_{\tilde{l}})$ as a function of $\Delta m_1/\Delta
  m_2 \equiv (m_{\tilde l}-m_{\chi_1^0}) / (m_{\chi_2^0}-m_{\chi_1^0})$ for
  three different values of $n \equiv m_{\chi_2^0}/m_{\chi_1^0}$.
}
 \label{fig:endb}
 \end{figure}

We define the enhancement factor by
$(\frac{\Delta m_{ll}}{m_{ll}})/(\frac{\Delta m_{\tilde{l}}}{m_{\tilde{l}}})$.
Fig.~\ref{fig:endb} shows the enhancement factor as a function of the slepton mass.
 As well as the mSUGRA bino-dominated lightest supersymmetric particle (LSP)
case $n \equiv m_{\chi_2^0}/m_{\chi_1^0}=2$ we also allow for a more
compressed spectrum with different values of $n$.
There are several notable points.
When $m_{\tilde{l}} =
\sqrt{m_{\chi_1^0}m_{\chi_2^0}}$ the shift in the edge vanishes to leading order in $\frac{\Delta m_{\tilde{l}}}{m_{\tilde{l}}}$.
Conversely, large splittings of the di-lepton and di-muon edges can be achieved
for relatively small splittings of the slepton and smuon masses.
The enhancement factor is larger for
more degenerate masses between sparticles in the chain and can easily be $\mc{O}(10)$
depending upon the value of $m_{\chi_2^0}/m_{\chi_1^0}$.
The enhancement diverges as the slepton mass approaches either neutralino mass.
The benefits of the enhancement may be diluted for highly degenerate spectra by
the fact that leptons coming
from such chains will tend to be softer and thus harder to identify and measure
experimentally,
 and also by any change in precision due to
rate changes from differing branching ratios.

The muon energy calibration will depend
heavily on their energies, but we may expect particularly accurate calibration
when
the $m_{ll}$ edge is close to, but not on, the $Z^0$ pole.
Electrons and muons from the $Z^0$ decays may then be used to calibrate
energies and efficiencies, performing a small extrapolation to the relevant
energy.
For instance, the scale of the $m_{ll}$ endpoint is 80, 118,
99, 122, 343 GeV for the Snowmass benchmark points SPS1a, 3, 5, 8 and 9
respectively~\cite{sps}. Thus 
4 of the 5 relevant endpoints are rather close in invariant mass to the $Z^0$.

\begin{figure}
\epsfig{file=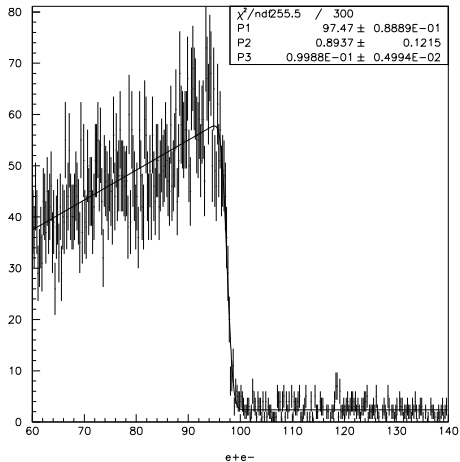, width=0.8 \columnwidth}
\caption{Log likelihood fit to di-electron edge in {\tt AcerDET} as
  described in the text. $m_{ee}$ has been placed in 0.25 GeV bins on
  abscissa. The ordinate shows events per bin per 16 $fb^{-1}$.}
 \label{fig:eeFit}
 \end{figure}

\begin{figure}
\epsfig{file=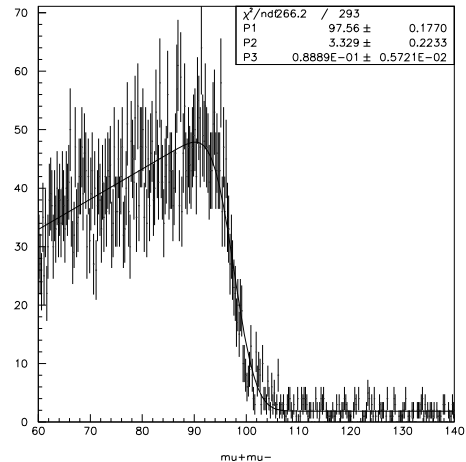, width=0.8 \columnwidth}
\caption{Log likelihood fit to di-muon edge in {\tt AcerDET} as described in the
  text. $m_{\mu\mu}$ has been placed in 0.25 GeV bins on 
  abscissa. The ordinate shows events per bin per 16 $fb^{-1}$. }
 \label{fig:mumuFit}
 \end{figure}

To estimate the experimental sensitivity to differences in the
positions of the di-muon and di-electron edge, we generate 400,000
$R$-parity conserving supersymmetric events at SUGRA Point 5 from
proton-proton collisions at a 14 TeV centre of mass energy.  We use
{\tt HERWIG 6.510} which reports the two-to-two (Standard Model to
Supersymmetric Particle) cross section to be $\sim 24~pb$.  Our sample
of 400,000 events therefore corresponds to an integrated luminosity of
$\sim 16~fb^{-1}$.  We pass the generated events through {\tt AcerDET}
\cite{RichterWas:2002ch} which represents a fast simulation of a
generic general-purpose LHC-type experiment.  The main feature that
{\tt AcerDET} provides for our purposes is a reasonable model of the
electron and muon momentum resolution for a typical LHC-type detector.
{\tt AcerDET} accomplishes this by `smearing' the generated momenta by
appropriate amounts dependent on $p_T$, $\eta$ and $\phi$.  {\tt
AcerDET} provides only minimal simulation of reconstruction
efficiencies.  For example, electrons will alway be reconstructed
unless they are out of acceptance or are too close to other particles.
Neither does {\tt AcerDET} model any uncertainty in the absolute energy
scale for either electrons or muons.  For this we 
use an estimate of 0.1\% \cite{Armstrong:1994it, CMSNote}.  Absolute energy scale calibration at this level might appear to be a tall order --  however the standard candles used in this calibration ($Z$-bosons) will be produced in such large numbers at the LHC that the estimate may even turn out to be conservative.  To evaluate the
sensitivity with which the endpoints of the di-electron and di-muon
invariant mass spectra can be measured, we generate their
distributions, fit them in the vicinity of the endpoint, and report
the endpoint fit uncertainties. To select our event we require two
opposite sign same family (OSSF) isolated leptons of greater than 10
GeV and missing energy greater than 100 GeV. At this point, the
di-lepton invariant mass distribution should in principle end at 97.48
GeV.  We fit over the range 60 GeV to 140 GeV.  For our (log
likelihood) fit we use a function which takes the form of a triangle
distribution (i.e. $f(x;e)\propto x \Theta(e-x)\Theta(x)$ where
$\Theta(x)$ is the Heaviside step-function and $e$ is the notional
endpoint) sitting on top of a constant background, and with the entire
distribution convolved with a Gaussian resolution of width $\sigma$.
In the fit, the three free parameters were (1) the endpoint position
$e$, (2) its resolution $\sigma$, and (3) the ratio of the
number of signal and SUSY background events.  The overall
normalisation was fixed analytically to allow a log-likelihood fit to
be performed with narrow bins (0.25 GeV).  Example fits are shown in
figures~\ref{fig:eeFit} and \ref{fig:mumuFit}.
Note that we neither generate
nor fit Standard Model (SM) events.  In reality, irreducible sources
of SM di-leptons from $Z$-bosons would constitute an important
background if the di-lepton endpoint were to occur on the $Z$ resonance.
We ignore this special case for the moment because it is our primary
aim to make statements about the ``generic'' di-lepton edge
sensitivity.  Away from the $Z$-peak, the missing energy cut will
reduce the SM backgrounds.  The results of the endpoint
fits are summarised in Table~\ref{tab:tab}.
 \begin{table}\begin{tabular}{|c|c|c|c|}
 \hline
 Integrated & Events & Electron & Muon \\
 Luminosity & below & Endpoint & Endpoint \\
 ($fb^{-1}$) & 100 GeV & (GeV) & (GeV) \\
 \hline
 16.0 &  22145 & $97.47 \pm  0.09$ & $97.56 \pm  0.18$ \\
  8.0 &  11131 & $97.41 \pm  0.13$ & $97.83 \pm  0.23$ \\
  4.0 &   5520 & $97.54 \pm  0.19$ & $97.63 \pm  0.35$ \\
  2.0 &   2707 & $97.52 \pm  0.28$& $97.56 \pm  0.50$ \\
 \hline
 \end{tabular}
 \caption{Results of the endpoint  fits for various integrated
 luminosities. The number of OSSF di-lepton events passing cuts
 and having di-lepton invariant mass below  100 GeV is also recorded.}
 \label{tab:tab}
 \end{table}

Accordingly we parametrise the LHC di-lepton edge
  sensitivity $\Sigma$ (which we define to be the expected fractional
  uncertainty in the di-muon di-electron edge difference as estimated
  by the fit error on the less well determined endpoint added in
  quadrature to the energy scale uncertainty) by
\begin{equation}
\Sigma = \sqrt{(0.002 \sqrt{22100 / N})^2 + 0.001 ^2},
\end{equation}
for $N$ expected signal events in the di-lepton channel.  In this
expression, the ``0.001 term'' represents the 0.1\% absolute energy
scale error described earlier.  Note that the  di-lepton edge
sensitivity $\Sigma$ is not to be confused with the slepton mass
sensitivity $E$ defined later.

By the above definition, the edge sensitivity
$\Sigma$ is a measure of the scale down to which fractional
differences in endpoint positions ${\Delta m_{ll}}/{m_{ll}}$ can be
measured.  More precisely, assuming that the endpoint fit error is
approximately Gaussian distributed, it should be possible to make an
``$S_1$-sigma'' discovery of selectron-smuon pole mass non-universality
(i.e.~rule out a null hypothesis of  ``no splitting in the di-electron
and di-muon endpoints'' at the $S_1$-sigma level) for a real endpoint
splitting of size $\Delta m_{ll}$ according to
\begin{equation}
  S_1 = \left| \frac{\Delta m_{ll}}{m_{ll}} \right| \div
\Sigma \label{signif}.
\end{equation}
In what follows we will therefore refer to $S_1$ as the ``discovery
significance'' for selectron-smuon mass non-universality.  We note
that, when calculating $S_1$, the numbers of events $N_{ee}$
contributing to the $e^+ e^-$ di-lepton signal may differ from the
number of $\mu^+ \mu^-$ pairs $N_{\mu\mu}$ due to phase-space
differences induced by the mass differences.  If systematic
uncertainties on trigger and reconstruction efficiencies can be
controlled, this could provide an additional means of testing
selectron-smuon mass universality by looking at significant
differences from zero in the statistic
\begin{equation}
S_2=\frac{N_{ee}-N_{\mu\mu}}{\sqrt{N}} \label{sig2}
\end{equation}
which, like $S_1$, will be approximately normally distributed.
We do not use $S_2$ ourselves.

\par

In the examples which follow we calculate the $N$ in $S_1$ for 30
fb$^{-1}$ of integrated luminosity at the LHC using {\tt
WIGISASUGRA1.200} and {\tt HERWIG6.5}~\cite{Corcella:2000bw}.
Performing a Markov Chain Monte Carlo maximisation on $S_1$ in mSUGRA,
we find a maximum value $S_1=0.52$ 
after direct search
constraints have been applied. Thus we find that the smuon-selectron
splitting cannot be discriminated from zero in mSUGRA at the LHC.  On
the other hand, any significant measured difference in the end points
at the LHC will discriminate against mSUGRA.  A future international
linear collider would achieve much improved accuracy~\cite{ILCLHC}
upon the mass splitting and could be combined with other constraints
to help bound $\tan \beta$ assuming mSUGRA.

We wish to emphasise that the reason mSUGRA fails to generate an
observable edge splitting is not the one often suggested.  It is
{\em not} true that the muon and electron Yukawa couplings are too small to
play any role in the RGEs. Indeed, at large $\tan \beta$ the RGEs,
combined with the enhancement factor, can generate slepton spectra
giving edge splittings at the per cent level.  The real reason is that
in this case
$\tilde{\tau}_R$ is driven light and it dominates the $\chi_2^0$ decay
modes, with $BR(\chi_2^0 \to \tilde{l}_R l) \ll 1$ and $BR(\chi_2^0 \to
\tilde{\tau}_1 \tau) \sim 1$.  In models where there is extra third
family physics that would lead to the weak scale mass ordering
$m_{\tilde{\tau}_R} > m_{\chi_2^0} > m_{\tilde{l}_R} > m_{\chi_1^0}$, the
selectron and smuon RGEs can be sufficient to generate electron-muon
edge splittings that can be significantly discriminated from zero.  As
an example, we consider the point $m_0=148$ GeV, $M_{1/2}=250$ GeV,
$A_0=-600$ GeV, $\tan \beta=40$ with $m_{{\tilde
\tau}_{L,R}}(M_{X})=950$ GeV.  At this point, $\Delta
m_{\tilde{l}}/m_{\tilde{l}}=2.3 \times 10^{-3}$ and $\Delta
m_{ll}/m_{ll}=1.5 \%$ whereas $\Sigma=0.27 \%$, allowing an $(S_1> 5)$-sigma
discovery significance for smuon-selectron pole mass non-universality.

\par

We now focus directly on the sensitivity to slepton mass splittings
and analyse how degenerate the selectron and smuon
masses can be while still allowing a 1-sigma sensitivity to a non-zero
mass difference.  We do this assuming 30fb$^{-1}$ in perturbed mSUGRA
around SPS1a. In perturbed mSUGRA, we take mSUGRA boundary conditions
but we allow $m_{{\tilde \mu}_R}$ to float away from the mSUGRA
prediction, as could be derived from $\Delta m_{\tilde l}^2 (M_X)\neq
0$ in Eq.~\ref{leadinglog}.  By using Eqs.~\ref{edgesplitting} and
\ref{signif}, setting $S_1$ to 1 
we obtain the fractional slepton mass
splitting which might be discriminated from zero at the 1-sigma level:
\begin{equation}
E \equiv
\left. \frac{{{\Delta m_{\tilde l}}}}{{ m_{\tilde l}}} \right|_{S_1=1} =  \frac{(m_{\chi_2^0}^2 - m_{\tilde{l}}^2 )
(m_{\tilde{l}}^2 - m_{\chi_1^0}^2)}{m_{\chi_1^0}^2
  m_{\chi_2^0}^2 - m_{\tilde{l}}^4} {\Sigma_{30 \rm {fb}^{-1}}},
\end{equation}
valid in the limit $\Delta m_{\tilde l} / m_{\tilde l} \ll 1$.
\begin{figure}
  \unitlength=1in
  \begin{picture}(4,2.38)
    \put(-0.4,-0.05){\epsfig{file=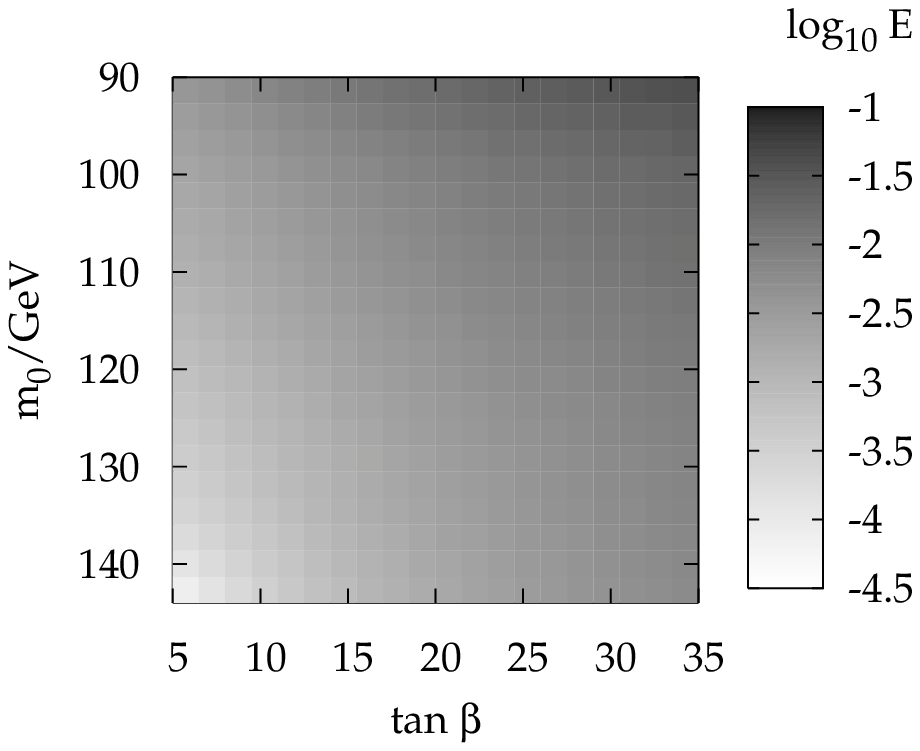, width=4in}}%
    \put(0.61,0.47){\epsfig{file=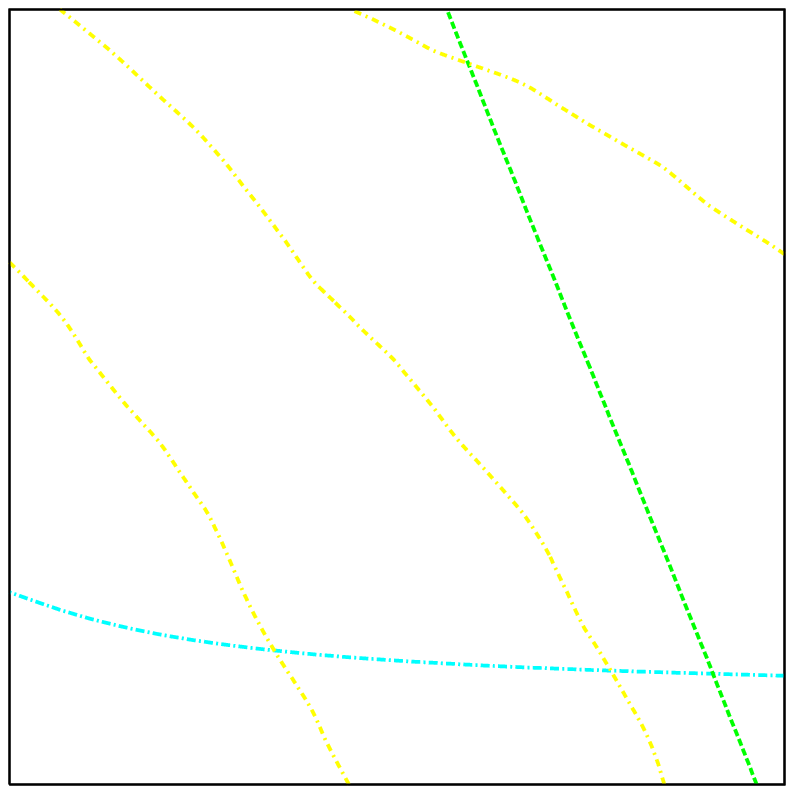, width=2.71in}}
  \end{picture}
  \caption{Expected 30fb$^{-1}$ 1-sigma sensitivity, $E$, to selectron-smuon
  mass splitting 
  in perturbed mSUGRA around SPS1a. The region to the right hand side of the
  almost-vertical line has $m_{{\tilde \tau}_1}<m_{\chi_1^0}$. The region
  underneath the mostly 
  horizontal line has $m_{\chi_2^0}-m_{\tilde l} < 10$ GeV.
The lighter lines
show contours of $\log_{10}E=-2,-2.5,-3$ (top to bottom)
}
  \label{fig:scanE}
\end{figure}
Fig.~\ref{fig:scanE} displays $E$ for a scan of perturbed mSUGRA around
SPS1a. 
The expected sensitivity at SPS1a itself is $E=2.8 \times 10^{-3}$.
The strict mSUGRA prediction
for the smuon-selectron mass splitting is $\Delta m_{\tilde l} / m_{\tilde l}
= 5.9 \times 10^{-5}$.
We see that sensitivities down
to $\mc{O}(10^{-4})$ are possible while restricting to the region where
$m_{\chi_2^0}-m_{\tilde l} > 10$ 
GeV to ensure sufficiently hard leptons.

\begin{figure}
  \vspace{0.2cm}
  \epsfig{file=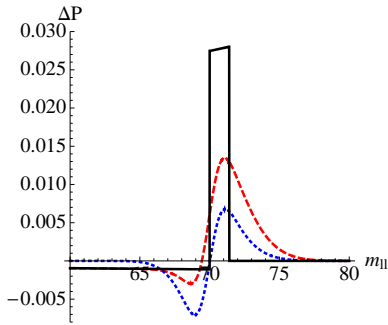, width=2in}%
  \caption{Probability distribution difference in GeV$^{-1}$ of SUSY cascade
    chain electrons and muons as a function of their invariant mass $m_{ll}$
    (in GeV) 
    for an endpoint of 70 GeV and a relative splitting of 2$\%$. The solid
    line shows the distribution without any energy resolution taken into
    account, whereas the dashed line displays the effect of smearing due to
    energy resolution. 
The dotted line shows a smeared distribution with 
    no mass splitting. 
  }
  \label{fig:dmll}
  \end{figure}

It is tempting to display 
the difference in $\mu^+\mu^-$ and $e^+e^-$
probability distributions ($\Delta P$) in order to look for a spike, see the
solid curve in 
Fig.~\ref{fig:dmll}. In practice, energy resolution effects would smear out
this curve. 
In the dashed
curve, $m_{\mu\mu}$ and $m_{ee}$ have been Gaussian
smeared by assumed fractional resolutions of 1$\%$ and
 3$\%$ respectively. We have assumed $N_{ee}=N_{\mu\mu}$ and simulated
no backgrounds.
We have also plotted $\Delta P$ for equal
endpoints while taking the energy resolution into account in the 
dotted line. Simply having a worse energy resolution for muons still leads to
a similar feature in the difference near the end-point and could mislead us
into 
thinking there is a splitting when in fact there is none. 
We conclude that, in practice, the best approach will be to fit
the di-electron and di-muon endpoints separately, and then examine the
difference, as was done in Figs.~\ref{fig:eeFit} and \ref{fig:mumuFit}.
We note here that one could still employ the technique of subtracting opposite
sign different flavor di-leptons from the di-muon or di-electron decay chain
samples in order to subtract backgrounds (e.g.\ from top pairs or $W$ pairs),
as was recommended in the  
case~\cite{Armstrong:1994it} where the two samples are summed. 

In this paper we have studied the sensitivity of the LHC to
the $\tilde \mu$-$\tilde e$ mass splitting through endpoint differences in SUSY
cascade decays. 
Enhancement factors mean that measured differences in the endpoints can be
a factor of ten more sensitive to tje mass splitting.
In the large $\tan \beta$ limit RGE can induce edge
splittings up to the per cent level. 
However, in mSUGRA the $\chi_2^0 \to \tilde{\tau}_1 \tau$ branching ratio
increases, significantly reducing the $\chi_2^0 \to \tilde{l}_R l$ branching
ratio. 
A significant rejection of the universal smuon-selectron pole mass hypothesis 
would discriminate against mSUGRA. 
If additional mass terms
were to cause $m_{{\tilde \tau}_{1,2}} > m_{\chi_2^0}$,
then the RGE-induced 
edge splitting may be significantly discriminated from zero.
The di-lepton edge splitting may be a powerful
measurement and discrimination tool in SUSY data analysis, and should not be
forgotten in global fits to data.

 \begin{acknowledgments}
 This work has been partially supported by STFC. We thank the other members of
 the  Cambridge SUSY Working Group for helpful suggestions and comments.
\end{acknowledgments}

\end{document}